\documentclass[twocolumn,secnumarabic,amssymb, nobibnotes, aps, prl]{revtex4-1}

\begin{document}

\title{Highly-damped quasi-normal frequencies for piecewise Eckart  potentials}%

\author{Jozef Skakala}%
\email{jozef.skakala@msor.vuw.ac.nz}
\affiliation{School of Mathematics, Statistics, and Operations Research; Victoria University of Wellington, Wellington, New Zealand.}
\author{Matt Visser}%
\email{matt.visser@msor.vuw.ac.nz}
\affiliation{School of Mathematics, Statistics, and Operations Research; Victoria University of Wellington, Wellington, New Zealand.}

\date{4 May 2010; 15 June 2010; \LaTeX-ed \today}%

\begin{abstract}
Highly-damped quasi-normal frequencies are very often of the form  $\omega_n =  \hbox{(offset)} + i n \; \hbox{(gap)}$. We investigate the genericity of this phenomenon by considering a model potential that is piecewise Eckart (piecewise P\"oschl--Teller), and developing an analytic ``quantization condition'' for the highly-damped quasi-normal frequencies. We find that this $\omega_n =  \hbox{(offset)} + i n \; \hbox{(gap)}$ behaviour is \emph{generic but not universal}, with the controlling feature being whether or not the ratio of the rates of exponential falloff in the two asymptotic directions is a rational number. These observations are of direct relevance  to any physical situation where highly-damped quasi-normal modes (damped modes) are important --- in particular (but not limited to) to black hole physics, both theoretical and observational.

Physical Review {\bf D81} (2010)125023.  doi:10.1103/PhysRevD.81.125023
\end{abstract}

\pacs{}

\maketitle
\def\sech{{\mathrm{sech}}}
\def\ln{{\mathrm{ln}}}
\def\d{{\mathrm{d}}}

In many diverse branches of physics one is interested in studying potentials that enjoy  suitable falloff conditions at spatial infinity, and it is generally observed that such potentials  lead to quasi-normal modes (QNMs, damped modes) with associated quasi-normal frequencies (QNFs). To help gain semi-analytic understanding of this phenomenon we investigate the QNFs of a piecewise Eckart (P\"oschl--Teller) potential~\cite{Eckart, Poeschl, Morse, Boonserm}. We are particularly interested in understanding the 
\begin{equation}
\omega_n =  \hbox{(offset)} + i n \; \hbox{(gap)}
\end{equation}
 behaviour that has been encountered in very many different analyses, often in the context of black hole physics, but by no means limited to black hole physics~\cite{Chandrasekhar, Kokkotas, Nollert, Berti, Dreyer, Natario, Ferrari, Iyer, Guinn1, Guinn2, Konoplya, Medved1, Medved2, Padmanabhan, Choudhury, Andersson1, Andersson2, Leaver1, Leaver2, Leaver3, Motl1, Motl2, Shanka1, Shanka2, Chan1, Chan2, Horowitz, Wang1, Wang2, Wang3, Cardoso, Wang4, Wang5, Wang6}.  

The specific model we are interested in  is
\begin{equation}
-\psi''(x) + V(x) \; \psi(x) = 0,
\end{equation}
with a piecewise Eckart potential~\cite{Jozef}
\begin{equation}
V(x) = \left\{  \begin{array}{lcl}
{V_{0-} \; \sech^2(x/b_-)}  & \hbox{ for }  & x < 0;\\
V_{0+} \; \sech^2(x/b_+)  & \hbox{ for }  & x > 0. \\
\end{array}\right.
\end{equation}
The standard case is  $V_{0-} = V_{0+} = V_0$, with $b_-=b_+ = b$, 
so that  $V(x) = {V_{0} \; \sech^2(x/b)}$.
A related model where $V_{0-} = V_{0+} = V_0$ but with $b_+\neq b_-$ has been explored by Suneeta~\cite{Suneeta}, but our current model is more general, and we will take the analysis much further. 
We start by imposing quasi-normal boundary conditions (outgoing radiation boundary conditions)~\cite{Jozef, Beyer}
\begin{equation}
\psi_+(x\to+\infty) \to e^{-i\omega x}; \quad \psi_-(x\to-\infty) \to e^{+i\omega x}.
\end{equation}
On each half line ($x<0$, and $x>0$) the exact wavefunction  (see especially page 405 of~\cite{Beyer}) is:
\begin{equation}
\!\!
\psi_\pm(x) = e^{\mp i\omega x} \; {} _2F_1\left({1\over2}+\alpha_\pm,{1\over2}-\alpha_\pm,1+ib_\pm \omega, z \right),
\end{equation}
where
\begin{equation}
\alpha = \left\{  \begin{array}{lcl}
\sqrt{{1\over4} - V_0 b^2 } & \hbox{ for }  & V_0 b^2 < 1/4;\\
i \sqrt{V_0 b^2 - {1\over4}  } & \hbox{ for }  & V_0 b^2 > 1/4;\\
\end{array}\right.
\end{equation}
and $z = 1/(1+ e^{\pm 2x/b_\pm})$.
The key step in matching these two exact wavefunctions at $x=0$ is to calculate the logarithmic derivative. 
Using the Leibnitz rule and the chain rule one evaluates ${\psi_\pm'(0)/\psi_\pm(0)}$ as:
\begin{equation}
\mp i \omega \mp    {1\over 2 b_\pm} \! \left. {\d \ln \left\{  {}_2F_1\left({1\over2}+\alpha_\pm,{1\over2}-\alpha_\pm,1+ib_\pm \omega,z\right) \right\} \over \d z}\right|_{z=1/2.} 
\end{equation}
Invoking the differential identity 
\begin{eqnarray}
\label{E:differential}
&& {\d \left\{ _2F_1\left(a,b,c,z\right) \right\} \over \d z} 
\nonumber
\\
&&
\qquad
= {c-1\over z} \left[  \; _2F_1\left(a,b,c-1,z\right) -  \;_2F_1\left(a,b,c,z\right) \right],
\end{eqnarray}
we see
\begin{equation}
{\psi_\pm'(0)\over\psi_\pm(0)} = \mp i \omega \; 
\left.
{_2 F_1\left({1\over2}+\alpha_\pm,{1\over2}-\alpha_\pm,ib_\pm \omega,z\right)
\over
_2F_1\left({1\over2}+\alpha_\pm,{1\over2}-\alpha_\pm,1+ib_\pm \omega,z\right)}
\right|_{z=1/2.} 
\end{equation}
Now using Bailey's theorem
\begin{equation}
\label{E:bailey}
_2F_1\left(a,1-a,c,{1\over2}\right) = {\Gamma({c\over2}) \Gamma({c+1\over2}) \over \Gamma({c+a\over2}) \Gamma({c-a+1\over2})},
\end{equation}
we have the exact result
\begin{equation}
{\psi_\pm'(0)\over\psi_\pm(0)} 
= \mp {2\over b_\pm} \; {\Gamma({\alpha_\pm+i\omega b_\pm\over2} +{3\over4}) \Gamma({-\alpha_\pm+i\omega b_\pm\over2} +{3\over4}) \over 
\Gamma({\alpha_\pm+i\omega b_\pm \over2} + {1\over4}) \Gamma({-\alpha_\pm+i\omega b_\pm \over2} + {1\over4})}.
\end{equation}
To obtain a more tractable result it is extremely useful to use the reflection formula 
\begin{equation}
\label{E:reflection}
\Gamma(z)\; \Gamma(1-z) = {\pi\over\sin(\pi z)},
\end{equation}
 to derive
\begin{eqnarray}
 {\Gamma({\alpha_\pm+i\omega b_\pm\over2} +{3\over4}) \over \Gamma({\alpha_\pm+i\omega b_\pm \over2} + {1\over4}) } 
&=&  { \Gamma(-{\alpha_\pm+i\omega b_\pm \over2} + {3\over4}) 
 \over
 \Gamma(-{\alpha_\pm+i\omega b_\pm\over2} +{1\over4})} 
 \nonumber\\
 && \times \tan\left(\pi\left[{\alpha_\pm+i\omega b_\pm\over2} +{1\over4}\right]\right). 
 \qquad
 \end{eqnarray}
This leads to the exact result
\begin{eqnarray}
{\psi_\pm'(0)\over\psi_\pm(0)} &=&  \mp {2\over b_\pm}  \;
 { \Gamma({-\alpha_\pm-i\omega b_\pm \over2} + {3\over4})  \Gamma({\alpha_\pm-i\omega b_\pm \over2} + {3\over4}) 
 \over
 \Gamma({-\alpha_\pm-i\omega b_\pm\over2} +{1\over4})  \Gamma({\alpha_\pm-i\omega b_\pm\over2} +{1\over4})} 
 \nonumber
\\
&&
 \times
   \; \tan\left(\pi\left[{\alpha_\pm+i\omega b_\pm\over2} +{1\over4}\right]\right) 
   \nonumber
\\
&&
 \times \; \tan\left(\pi\left[{-\alpha_\pm+i\omega b_\pm\over2} +{1\over4}\right]\right).
 \end{eqnarray}
If $\omega$ has a large positive imaginary part, then the Gamma function arguments above tend towards the positive real axis, a region where the Gamma function is smooth --- all potential poles in the logarithmic derivative have been isolated in the trigonometric functions. 
We now use the trigonometric  identity 
 \begin{equation}
 \label{E:trig1}
 \tan A \; \tan B = { \cos(A-B) - \cos(A+B)\over\cos(A-B) + \cos(A+B)};
 \end{equation}
 to rewrite this as
  \begin{eqnarray}
{\psi_\pm'(0)\over\psi_\pm(0)} &=&  \mp {2\over b_\pm}  \;
 { \Gamma({-\alpha_\pm-i\omega b_\pm \over2} + {3\over4})  \Gamma({\alpha_\pm-i\omega b_\pm \over2} + {3\over4}) 
 \over
 \Gamma({-\alpha_\pm-i\omega b_\pm\over2} +{1\over4})  \Gamma({\alpha_\pm-i\omega b_\pm\over2} +{1\over4})} 
 \nonumber\\
 &&
 \times  {\cos(\pi\alpha_\pm) + \sin(i\pi\omega b_\pm) \over \cos(\pi\alpha_\pm) - \sin(i\pi\omega b_\pm) }.
 \end{eqnarray}
The exact junction condition we wish to apply is
 \begin{equation}
{\psi_+'(0)\over\psi_+(0)}  = {\psi_-'(0)\over\psi_-(0)}.
\end{equation} 
As long as we are primarily focussed on the highly damped QNFs ($Im(\omega)\to\infty$) we can employ Stirling's approximation to deduce
\begin{equation}
\label{E:stirling}
 {\Gamma(z+{1\over2})\over\Gamma(z)} = \sqrt{z} \;  \left[1+ O\left({1\over z}\right)\right];   \qquad Re(z)\to \infty.
\end{equation}
Therefore 
\begin{eqnarray}
 { \Gamma({\pm\alpha_\pm-i\omega b_\pm \over2} + {3\over4})  
 \over
  \Gamma({\pm\alpha_\pm-i\omega b_\pm\over2} +{1\over4})  }
  &=& \sqrt{ {Im(\omega) b_\pm\over 2} } \; \left[1+ O\left({1\over Im(\omega b_\pm)}\right)\right].
  \nonumber\\
  &&
  \end{eqnarray}
This allows us to deduce an \emph{approximate} junction condition for the asymptotic QNFs 
\begin{equation}
 {\cos(\pi\alpha_+) + \sin(i\pi\omega b_+) \over \cos(\pi\alpha_+) - \sin(i\pi\omega b_+) } = 
 -  {\cos(\pi\alpha_-) + \sin(i\pi\omega b_-) \over \cos(\pi\alpha_-) - \sin(i\pi\omega b_-) },
\end{equation}
which is accurate up to fractional corrections of order $O\left({1/ Im(\omega b_\pm)}\right)$.
This asymptotic QNF condition can be rewritten in any one of the equivalent forms:
 \begin{equation}
 \label{E:qnf1}
\sin(-i\pi\omega b_+)\sin(-i\pi\omega b_-)  =  \cos(\pi\alpha_+)\cos(\pi\alpha_-);
 \end{equation}
  \begin{equation}
   \label{E:qnf2}
\sinh(\pi\omega b_+)\sinh(\pi\omega b_-)  =  - \cos(\pi\alpha_+)\cos(\pi\alpha_-);
 \end{equation}
\begin{eqnarray}
 \label{E:qnf3}
\cos(-i\pi\omega[b_+ -b_-]) - \cos(-i\pi\omega[b_+ + b_-]) 
\nonumber
\\
= 2  \;  \cos(\pi\alpha_+)\cos(\pi\alpha_-);
 \end{eqnarray}
 \begin{eqnarray}
 \label{E:qnf4}
\cosh(\pi\omega[b_+ -b_-]) - \cosh(\pi\omega[b_+ + b_-]) 
\nonumber\\
= 2  \;  \cos(\pi\alpha_+)\cos(\pi\alpha_-).
 \end{eqnarray}
Now suppose $b_+/b_-$ is rational, that is
\begin{equation}
{b_+/ b_-} = {p_+/ p_-},
\end{equation}
and suppose we define $b_*$ by
\begin{equation}
b_+= p_+ b_*; \qquad b_- = p_- b_*; \qquad b_* = \mathrm{hcf}(b_+, b_-),
\end{equation}
then the asymptotic QNF condition is given by
 \begin{equation}
 \sin(-i\omega \pi p_+ b_*)\sin(-i\omega \pi p_- b_*)  =  \cos(\pi\alpha_+)\cos(\pi\alpha_-).
 \end{equation}
If $\omega_*$ is \emph{any} specific solution of this equation, then
\begin{equation}
\omega_n = \omega_* + {in\over b_*} = \omega_* + in \; \mathrm{lcm}\left({1\over b_+}, {1\over b_-}\right)
\end{equation}
will also be a solution.
To characterize \emph{all} the solutions, consider the set of QNFs for which 
\begin{equation}
Im(\omega) < {1/ b_*},
\end{equation}
and label them as
\begin{equation}
\omega_{0,a} \qquad a\in\{1,2,3\dots N\}.
\end{equation}
Then the set of QNFs decomposes into a set of ``families''
\begin{equation}
\omega_{n,a} = \omega_{0,a} + {in/ b_*}; 
\end{equation}
with $a\in\{1,2,3\dots N\}$ and $ n\in\{0,1,2,3\dots\}$, and 
where $N$ is yet to be detemined. 
But for rational  $b_+/b_-$  we can rewrite the QNF condition as
\begin{eqnarray}
\cos(-i\omega\pi b_*|p_+-p_-|) - \cos(-i\omega\pi b_*[p_+ +p_-])  
\nonumber\\
= 2\cos\left(\pi\alpha_+\right) \cos\left(\pi\alpha_-\right).
\end{eqnarray}
Now define $z=\exp(\omega\pi b_*)$,  then the QNF condition is
\begin{eqnarray}
z^{|p_+-p_-|} + z^{-|p_+-p_-|} - z^{[p_+ +p_-]} - z^{-[p_+ +p_-]} 
\nonumber\\
=  4\cos\left(\pi\alpha_+\right) \cos\left(\pi\alpha_-\right).
\end{eqnarray}
Equivalently
\begin{eqnarray}
z^{2[p_+ +p_-]} - z^{2p_\mathrm{max}} + 4\cos\left(\pi\alpha_+\right) \cos\left(\pi\alpha_-\right) z^{+[p_+ +p_-]}
\nonumber
\\
 -  z^{2p_\mathrm{min}} + 1 = 0.\qquad
\end{eqnarray}
This is a polynomial of degree $N=2(p_++p_-)$, so it has exactly $N$ roots $z_a$ (occurring in complex conjugate pairs). Thus the QNFs are
 \begin{equation}
 \omega_{n,a} = 
  { \ln(z_a) \over \pi b_*} + {in\over b_*},
 \end{equation}  
 with the imaginary part of the logarithm lying in $[0,2\pi)$, and
where $ a\in\{1,2,3\dots N\}$ and $n\in\{0,1,2,3,\dots\}$. 

 So for rational $b_+/b_-$ with $b_+/b_- = p_+/p_-$ we have exactly $N=2(p_++p_-)$ families of  equi-spaced QNFs all with with gap $i/b_*$ and with (typically distinct) offsets $  { \ln(z_a) /( \pi b_*) } $. 
 That is: \emph{Arbitrary rational ratios of ${b_+/ b_-} $ automatically  imply the}  $\omega_n =  \hbox{(offset)} + i n \; \hbox{(gap)}$ \emph{behaviour.}

Now in contrast suppose $b_+/b_-$ is irrational, that is
\begin{equation}
b_*=  \mathrm{hcf}(b_+, b_-) = 0.
\end{equation}
Then the ``families'' each have only  one element
\begin{equation}
\omega_{0,a}  \qquad a\in\{1,2,3\dots \infty\}. 
\end{equation}
That is, there will be no ``pattern'' in the QNFs, and they will not be regularly spaced. (Conversely, if there is a ``pattern'' then  $b_+/b_-$ is rational.)
Stated more formally: 
Suppose we have at least one family of equi-spaced QNFs such that
\begin{equation}
\label{E:equal}
\omega_n = \omega_0 + i n K,
\end{equation}
then  $b_+/b_-$ is rational.

To see this: If we have a family of QNFs of the form given in equation (\ref{E:equal})
then we know that $\forall n \geq 0$
\begin{eqnarray}
&&
+\cos(-i\omega_0\pi |b_+-b_-| + nK\pi |b_+-b_-| ) 
\nonumber\\
&&- \cos(-i\omega_0\pi [b_+ +b_-] + nK\pi |b_++b_-| ) \qquad
\nonumber\\
&& \qquad
= \cos(-i\omega_0\pi |b_+-b_-| ) - \cos(-i\omega_0\pi [b_+ +b_-]  ).\qquad
\end{eqnarray}
Let us write this in the form $\forall n \geq 0$
\begin{equation}
\cos(A+n J) - \cos(B+nL) = \cos(A)-\cos(B),
\end{equation}
and realize that this also implies
\begin{equation}
\label{E:1}
\cos(A+[n+1] J) - \cos(B+[n+1]L) = \cos(A)-\cos(B),
\end{equation}
and
\begin{equation}
\cos(A+[n+2] J) - \cos(B+[n+2]L) = \cos(A)-\cos(B). 
\end{equation}
Now appeal to the trigonometric identity 
\begin{equation}
\cos(A+[n+2] J) + \cos(A+n J) = 2 \cos(J) \cos(A+[n+1] J),
\end{equation}
to deduce
\begin{eqnarray}
\label{E:2}
\cos(J) \cos(A+[n+1] J) - \cos(L) \cos(B+[n+1]L) 
\nonumber\\
= \cos(A)-\cos(B).
\qquad
\end{eqnarray}
That is, $\forall n\geq 0$ we have \emph{both} (\ref{E:1}) and (\ref{E:2}). 
The first of these equations asserts that all the points 
\begin{equation}
\left( \vphantom{\Big|} \cos(A+[n+1] J) , \; \cos(B+[n+1]L)  \right)
\end{equation} 
lie on the straight line of slope 1 that passes through the point $(0, \cos B-\cos A)$. 
The second of these equations asserts that these same points 
\emph{also} lie on the straight line of slope $\cos(J)/\cos(L)$ that passes through the point $(0, [\cos B-\cos A]/\cos L)$. We then argue as follows:

\noindent $|$i) If $\cos J \neq \cos L$ then these two lines are not parallel and so meet only at a single point, let us call it $(\cos A_*, \cos B_*)$, whence we deduce
\begin{equation}
\cos(A+[n+1] J) = \cos A_* ; \;\;\; \cos(B+[n+1]L) = \cos B_*.
\end{equation}
But then both $J$ and $L$ must be multiples of $2\pi$, and so $\cos J = 1=  \cos L$ contrary to hypothesis.

\noindent $|$ii) If  $\cos J = \cos L \neq 1$ then we have both
\begin{equation}
\cos(A+[n+1] J) - \cos(B+[n+1]L) = \cos(A)-\cos(B),
\end{equation}
and
\begin{eqnarray}
\cos(J)\left[  \cos(A+[n+1] J) -  \cos(B+[n+1]L) \right] = 
\nonumber\\
\cos(A)-\cos(B).\qquad
\end{eqnarray}
but these are two parallel lines, both of slope 1, that never intersect unless $\cos(J)=1$.
Thus  $\cos J = 1=  \cos L$ contrary to hypothesis.

\noindent $|$iii)
We therefore conclude that both $J$ and $L$ must be multiples of $2\pi$, so that in particular $\cos J = 1=  \cos L$ (in which case the QNF condition is certainly satisfied).

But now
\begin{equation}
{|b_+-b_-|/( b_++b_-)} = {J/ L} \; \in\; \mathbf{Q},
\end{equation}
and therefore 
\begin{equation}
{b_+ / b_-}  \;\in\; \mathbf{Q}.
\end{equation}
That is: \emph{Rational ratios of ${b_+/ b_-} $ are implied by the}  $\omega_n =  \hbox{(offset)} + i n \; \hbox{(gap)}$ \emph{behaviour}. 
Thus we have demonstrated  that the   $\omega_n =  \hbox{(offset)} + i n \; \hbox{(gap)}$ behaviour is \emph{generic but not universal}, and is intimately related to the rationality (or otherwise) of the ratio of the $e$-folding parameters $b_\pm$. 

We have also checked that the analysis sketched above satisfies several appropriate consistency checks and has suitable well-behaved limits~\cite{Jozef}. A particularly important case (not dealt with in~\cite{Jozef}) is to consider the situation where one side of the potential exhibits power law (rather than exponential) falloff.  For example, for black hole physics a particularly common situation is 
\begin{equation}
V(x) \to V_{0+} \; {a^2\over (x+a)^2} \quad \hbox{for} \quad x\to+\infty.
\end{equation}
The exact wavefunctions for this potential can be written down in terms of Bessel functions, and in the limit of highly damped QNFs one can easily see
\begin{equation}
{\psi_+'(0)\over\psi_+(0)} \to  Im(\omega),
\end{equation}
leading to the very simple asymptotic QNF condition
\begin{equation}
1 = 
  {\cos(\pi\alpha_-) + \sin(i\pi\omega b_-) \over \cos(\pi\alpha_-) - \sin(i\pi\omega b_-) },
\end{equation}
whence, in this particular situation with a one-sided exponential falloff one asymptotically has
\begin{equation}
\omega_n = {in\over b_-}.
\end{equation}
This observation is useful in that it indicates that one-sided exponential falloff can be treated via a minor variant of the analysis in~\cite{Jozef}, and a power law falloff in the potential exhibits behaviour qualitatively similar to the limit $b_+\to\infty$. (As it should on physical grounds.)

Turning to specific applications in black hole physics: Will the general $\omega_n =  \hbox{(offset)} + i n \; \hbox{(gap)}$ behaviour discussed above extend to more ``realistic'' astrophysical or de Sitter black holes? Consider a ``wavepacket'' centered near the peak of the Regge--Wheeler (Zerelli) potential that is  built up out of  highly damped modes. While the initial short-time behaviour of the wavepacket is likely to be sensitive to the details of the Regee--Wheeler (Zerelli) potential, such a wavepacket will quickly damp out and spread out towards both $r_*\to-\infty$ and $r_*\to +\infty$, so that the wavepacket will penetrate regions where our piecewise Eckart model potential should be a good approximation to the true potential. We should therefore expect the results of our semi-analytic model to be qualitatively (but not necessarily quantitatively) accurate for estimating the asymptotic QNFs of ``realistic'' black holes. Because of the way the asymptotic QNF condition was derived, we do not expect our model to give good results for low-lying QNFs. 

 Of course one of the key points here is that most of the analysis is largely independent of black hole physics, and depends only on the falloff conditions placed on the potential near asymptotic infinity --- indeed if we completely forget the black hole motivation, it is already of considerable mathematical and physical interest that we have a nontrivial extension of the Eckart potential for which the QNFs are asymptotically exactly solvable --- one could in principle loop back to Eckart's original article and start asking questions about tunelling probabilities for electrons encountering such piecewise Eckart barriers.



\end{document}